\documentstyle[12pt]{article}
\input amssym.def

\def\rset{\Bbb{R}}

\def\rmax{\rset_{\max}}
\def\rmin{\rset_{\min}}

\def\pcle{\preceq}
\def\Mat{\rm{Mat}}
\def\0{\bf{0}}
\def\1{\bf{1}}

\title{Unifying Approach to Software and
Hardware Design for Scientific Calculations}

\author{G. L. Litvinov, V. P. Maslov, A. Ya. Rodionov
\thanks{The work was supported by the joint INTAS--RFBR
grant N 95--91.}}
\date{}

\begin{document}
\maketitle

\begin{abstract}
A unifying approach to software and hardware design generated by
ideas of Idempotent Mathematics is discussed. The so-called
idempotent correspondence
principle for algorithms, programs and hardware units is described.
A software project based on this approach is presented.

\end{abstract}

\section{Introduction}

     Numerical computations are   still  very  important  in  computer
applications.  But until recently there  was  a discrepancy
between   numerical   methods and
software/hardware   tools   for   scientific calculations.
In   particular, numerical programming  was  not  much influenced by the
progress in Mathematics, programming  languages  and  technology.
Modern tools for numerical calculations are not unified, standardized
and reliable enough. It is difficult to ensure
the necessary accuracy and safety of calculations without loss of the
efficiency and speed of data processing. It  is  difficult  to  get
correct and exact estimations of calculation errors. For example,
standard methods of interval arithmetic \cite{2} do not  allow  to
take  into account the error auto-correction effects
\cite{19}  and,  as   a  result,  to
estimate calculation errors accurately.

However, new ideas in Mathematics and Computer Science lead to a very
promising approach (initially presented in \cite{20}--\cite{22}).
 An essential aspect of this approach is developing a system
of algorithms, utilities and programs based on a new mathematical
calculus which is called {\it Idempotent Analysis}
or {\it Idempotent Calculus}, or {\it Idempotent Mathematics} etc.
For  many  problems  in  optimization  and
mathematical modeling this Idempotent Calculus plays the same unifying
role as Functional Analysis in Mathematical Physics, see, e.g., \cite{14},
\cite{17}, \cite{28}--\cite{30} and surveys \cite{15}, \cite{21}.

Idempotent Analysis  is  based  on replacing the usual arithmetic
operations by a new set  of basic operations (such as maximum
or minimum). There are a lot of such new arithmetics which are associated
with sufficiently  rich   algebraic   structures   called   idempotent
semirings. It  is very important that many problems,  nonlinear in the
usual sense,  become  linear  with  respect  to  an  appropriate   new
arithmetic, i.e.  linear  over a suitable semiring (the so-called
{\it idempotent superposition principle} \cite{26}, \cite{27}, \cite{17},
which is a natural analog of the well-known superposition
principle in Quantum Mechanics). This `linearity'
considerably simplifies explicit  constructions  of  their  solutions.
Examples are   the  Bellman  equation  and  its  generalizations,  the
Hamilton--Jacobi equation etc. The idempotent analysis is a powerful
heuristic tool to construct new algorithms and apply unexpected
analogies and ideas borrowed, e.g., from mathematical physics and
quantum mechanics.
 The  abstract theory is well advanced and includes,
in particular, a new integration  theory,  linear  algebra
and  spectral
theory, idempotent functional analysis,  idempotent Fourier transforms
and so on. Its applications include various optimization problems
such as multi-criteria decision making, optimization on graphs,
discrete optimization with a large parameter (asymptotic problems),
optimal design of computer systems and computer media, optimal
organization of parallel data processing, dynamic programming,
applications to differential equations, numerical analysis,
discrete event systems, computer science, discrete mathematics,
mathematical logic, etc. (see, e.g. \cite{1}, \cite{3},
\cite{5}--\cite{15},
\cite{17}, \cite{21}, \cite{26}--\cite{32}
and references therein).

It is possible to obtain an implementation of the new approach
to scientific and numeric calculations in the form of a powerful
software system
based on unified algorithms. This approach ensures the arbitrary
necessary accuracy and safety of numerical calculations and
a working time
reduction for programmers and users because of a software unification.

Our approach uses the techniques of object oriented and functional
programming (see, for example, \cite{25}, \cite{16}) which is very
convenient for the design of our (suggested) software system.
A computer algebra techniques \cite{4} is also used. The modern
techniques of systolic processors and VLSI realizations of
numerical algorithms including parallel algorithms of linear algebra
(see, for example, \cite{18}, \cite{31}) is convenient for effective
implementations of the proposed approach to hardware design.

There is a  regular  method  based  on  the theory for constructing
back-end processors and technical devices intended for a
realization of basic algorithms of idempotent calculus and
mathematics of semirings. These hardware facilities can increase the
speed of data processing.

\section{Mathematical objects and their computer representations}

     Numerical algorithms  are  combinations  of basic operations.
Usually these basic operations deal with `numbers'.  In fact these
`numbers' are  thought  of  as  members  of some numerical {\it domains}
(real numbers, integers etc.). But every computer calculation
deals with finite {\it models} (or finite {\it computer representations})
 of these numerical
domains. For example, integers can be modeled by integers modulo a
power of  number  2,  real  numbers  can be represented
by rational numbers or floating-point numbers etc. Discrepancies
between mathematical    objects (e.g. `ideal' numbers) and their
computer models
(representations) lead to calculation errors.

Due to imprecision of sources of input data in real-world problems, the
data usually come in the form of confidence intervals or other
number sets
rather than exact quantities. Interval Analysis (see, e.g.,~\cite{2})
extends operations of traditional calculus from numbers to
number intervals to make possible processing such imprecise data and
controlling rounding errors in computational mathematics.

     However, there are no universal models which are good  in  all
cases and  we  have  to  use  varieties  of  computer models.  For
example, real numbers can be represented by the following computer
models:

     standard floating-point numbers,

     double precision floating-point numbers,

     arbitrary precision floating-point numbers,

     rational numbers,

     finite precision rational numbers,

     floating-slash and fixed-slash rational numbers,

     interval numbers, etc.

To examine an algorithm it is often useful to have a possibility
to change computer representations of input/output data. For this aim
the corresponding algorithm (and its software implementation)
must be universal enough.

\section{Universal algorithms}

     It is very important that many algorithms do not depend on
particular models of a numerical domain and even on this domain
itself. Algorithms of linear algebra (matrix multiplication, Gauss
elimination etc.) are good examples of algorithms of this
type.

Of course, one algorithm may be more universal than another algorithm of
the same type. For example, numerical integration algorithms based on
the Gauss--Jacobi quadrature formulas actually depend on computer
models because they use finite precision constants. On the contrary,
the rectangular formula and the trapezoid rule do not depend
on models and in principle can be used even in the case of
idempotent integration (see below).

The so-called object oriented software tools and
programming languages (like $C^{++}$ and Java, see, e.g., \cite{25})
are very convenient for computer implementation of universal
algorithms.

     In fact there are  no  reasons  to  restrict  ourselves  with
numerical domains only. Actually it may be a ring of polynomials, a
field of rational functions,  or an idempotent semiring. The case
of idempotent semirings
is extremely   important   because  of  numerous  applications.

\section{Idempotent correspondence principle}

There is a nontrivial analogy between Mathematics of semirings and
Quantum Mechanics. For example, the field of real numbers can be
treated as a `quantum object' with respect to idempotent semirings.
So idempotent semirings can be treated as `classical' or
`semi-classical' objects with respect to the field of real
numbers.

Let $\rset$ be the field of real numbers and $\rset_+$ the subset
of all non-negative numbers. Consider the following change of variables:
$$
u \mapsto w = h \ln u,
$$
where $u \in \rset_+ \setminus \{0\}$, $h > 0$; thus $u = e^{w/h}$,
$w \in
\rset$. Denote by $\0$ the additional element $-\infty$ and by $S$
the extended real line $\rset \cup \{\0\}$. The above change of
variables has a
natural extension $D_h$ to the whole $S$ by $D_h(0) = \0$; also, we
denote $D_h(1) = 0 = \1$.

Denote by $S_h$ the set $S$ equipped with the two operations $\oplus_h$
(generalized addition) and $\odot_h$ (generalized multiplication)
such that
$D_h$ is a homomorphism of $\{\rset_+, +, \cdot\}$ to $\{S, \oplus_h,
\odot_h\}$. This means that $D_h(u_1 + u_2) = D_h(u_1) \oplus_h D_h(u_2)$
and $D_h(u_1 \cdot u_2) = D_h(u_1) \odot_h D_h(u_2)$, i.e., $w_1 \odot_h
w_2 = w_1 + w_2$ and $w_1 \oplus_h w_2 = h \ln (e^{w_1/h} + e^{w_2/h})$.
It is easy to prove that $w_1 \oplus_h w_2 \to \max\{w_1, w_2\}$
as $h \to 0$.

Denote by $\rmax$ the set $S = \rset \cup \{\0\}$ equipped
with operations
$\oplus = \max$ and $\odot = +$, where ${\0} = -\infty$, ${\1} = 0$
as above.
Algebraic structures in $\rset_+$ and $S_h$ are isomorphic; therefore
$\rmax$ is a result of a deformation of the structure in $\rset_+$.

We stress the obvious analogy with the quantization procedure,
where $h$ is
the analog of the Planck constant. In these terms, $\rset_+$
(or $\rset$)
plays the part of a `quantum object' while $\rmax$ acts as a
`classical' or `semi-classical' object that arises as the result
of a {\it dequantization} of this quantum object.

Likewise, denote by $\rmin$ the set $\rset \cup \{\0\}$ equipped with
operations $\oplus = \min$ and $\odot = +$, where ${\0} = +\infty$ and
${\1} = 0$. Clearly, the corresponding dequantization procedure is
generated by the change of variables $u \mapsto w = -h \ln u$.

Consider also the set $\rset \cup \{\0, \1\}$, where ${\0} = -\infty$,
${\1} =+\infty$, together with the operations $\oplus = \max$
and $\odot=\min$.
Obviously, it can be obtained as a result of a `second dequantization'
of $\rset$ or $\rset_+$.

The algebras presented in this section are the most important
examples of idempotent semirings,
the central algebraic structure of Idempotent Analysis.

Consider a set $S$ equipped with two algebraic operations: {\it addition}
$\oplus$ and {\it multiplication} $\odot$. The triple $\{S, \oplus,
\odot\}$ is an {\it idempotent semiring} if it
satisfies the following conditions (here and below, the symbol $\star$
denotes any of the two operations $\oplus$, $\odot$):
\begin{itemize}
\item the addition $\oplus$ and the multiplication $\odot$ are
associative: $ x \star (y \star z) = (x \star y) \star z$ for all
$x, y, z\in S$;
\item the addition $\oplus$ is commutative: $x \oplus y = y \oplus
x$ for all $x,y \in S$;
\item the addition $\oplus$ is {\it idempotent}: $x \oplus x = x$
       for all $x\in S$;
\item the multiplication $\odot$ is {\it distributive} with respect to
the
addition $\oplus$: $x\odot(y\oplus z) = (x\odot y)\oplus(x\odot z)$ and
$(x\oplus y)\odot z = (x\odot z)\oplus(y\odot z)$ for all $x, y, z\in
S$.
\end{itemize}

A {\it unity} of a semiring $S$ is an element ${\1}\in S$ such that
for all
$x \in S$
$$
   {\1} \odot x = x \odot {\1} = x.
$$

A {\it zero} of a semiring $S$ is an element ${\0} \in S$ such
that $\0 \neq
\1$ and for all $x \in S$
$$
 {\0}\oplus x = x,\qquad {\0}\odot x = x\odot {\0} = {\0}.
$$

A semiring $S$ is said to be {\it commutative} if $x\odot y=y\odot x$
for all $x,y\in S$. A commutative semiring is called a {\it
semifield} if every nonzero element of this semiring is invertible.
It is clear that $\Bbb R_{\max}$ and $\Bbb R_{\min}$ are semifields.

Note that different versions of this axiomatics are used, see, e.g.,
\cite{1}, \cite{3}, \cite{5}, \cite{6}, \cite{12}, \cite{13}--\cite{15},
\cite{17}, \cite{21}, \cite{23}, \cite{30} and some literature
indicated in these books and papers.
Many nontrivial examples of idempotent semirings can be found, e.g.,
in \cite{1}, \cite{5}, \cite{6}, \cite{12}, \cite{13}, \cite{14},
\cite{17}, \cite{21}, \cite{23}, \cite{24}, \cite{30}. For example,
every vector lattice or ordered group can be treated as an idempotent
semifield.

The addition $\oplus$ defines the following {\it standard partial order}
on $S$: $x\pcle y$ if and only if $x\oplus y=y$. If $S$ contains a zero
$\0$, then ${\0}\pcle x$ for all $x\in S$. The operations $\oplus$ and
$\odot$ are consistent with this order in the following sense: if
$x\pcle y$, then $x\star z\pcle y\star z$ and $z\star x\pcle z\star y$
for all $x,y,z \in S$.

The basic object of the traditional calculus is a {\it function}
defined on some set $X$ and taking its values in the field $\rset$
(or $\Bbb{C}$); its idempotent analog is a map $X \to S$, where $X$ is
some set and $S =\rmin$, $\rmax$, or another idempotent semiring. Let us
show that redefinition of basic constructions of traditional calculus in
terms of Idempotent Mathematics can yield interesting and nontrivial
results
(see, e.g., \cite{17}, \cite{21}, \cite{23}, \cite{24}, for details,
additional examples and generalizations).

{\sc Example 1. Integration and measures.} To define an idempotent
analog of the Riemann integral, consider a Riemann sum for a function
$\varphi(x)$, $x \in X = [a,b]$, and substitute semiring
operations $\oplus$
and $\odot$ for operations $+$ and $\cdot$ (usual addition and
multiplication) in its expression (for the sake of being definite,
consider the semiring $\rmax$):
$$
\sum_{i = 1}^N \varphi(x_i) \cdot \Delta_i \quad\mapsto\quad
\bigoplus_{i = 1}^N \varphi(x_i) \odot \Delta_i
= \max_{i = 1, \ldots, N}\, (\varphi(x_i) + \Delta_i),
$$
where $a = x_0 < x_1 < \cdots < x_N = b$, $\Delta_i = x_i - x_{i - 1}$, $i
= 1,\ldots,N$. As $\max_i \Delta_i \to 0$, the integral sum tends to
$$
 \int_X^\oplus \varphi(x)\, dx = \sup_{x \in X} \varphi(x)
$$
for any function $\varphi$:~$X \to \rmax$ that is bounded. In general,
for any set $X$ the set function
$$
m_\varphi(B) = \sup_{x \in B} \varphi(x), \quad B \subset X,
$$
is called an $\rmax$-{\it measure} on $X$;
since $m_\varphi(\bigcup_\alpha
B_\alpha) = \sup_\alpha m_\varphi(B_\alpha)$, this measure is completely
additive. An idempotent integral with respect to this measure
is defined as
$$
 \int_X^\oplus \psi(x)\, dm_\varphi
 = \int_X^\oplus \psi(x) \odot \varphi(x)\, dx
 = \sup_{x \in X}\, (\psi(x) + \varphi(x)).
$$

Using the standard partial order it is possible to generalize these
definitions for the case of arbitrary idempotent semirings.

{\sc Example 2. Fourier--Legendre transform.} Consider the topological
group $G = \rset^n$. The usual Fourier--Laplace transform is defined as
$$
\varphi(x) \mapsto \widetilde\varphi(\xi)
 = \int_G e^{i\xi \cdot x} \varphi(x)\, dx,
$$
where $\exp(i\xi \cdot x)$ is a {\it character} of the group $G$, i.e.,
a solution of the following functional equation:
$$
   f(x + y) = f(x)f(y).
$$

The idempotent analog of this equation is
$$
 f(x + y) = f(x) \odot f(y) = f(x) + f(y).
$$
Hence `idempotent characters' of the group $G$ are linear
functions of the
form $x \mapsto \xi \cdot x = \xi_1 x_1 + \cdots + \xi_n x_n$. Thus
the Fourier--Laplace transform turns into
$$
 \varphi(x) \mapsto \widetilde\varphi(\xi)
 = \int_G^\oplus \xi \cdot x \odot \varphi(x)\, dx
 = \sup_{x \in G}\, (\xi \cdot x + \varphi(x)).
$$
This is the well-known Legendre (or Fenchel) transform.

These examples suggest the following formulation of the idempotent
correspondence principle \cite{20}, \cite{21}:
\begin{quote}
{\it There exists a heuristic correspondence between interesting, useful,
and important constructions and results over the field of real (or
complex) numbers and similar constructions and results over idempotent
semirings in the spirit of N. Bohr's correspondence principle in
Quantum Mechanics.}
\end{quote}

So Idempotent Mathematics can be treated as a `classical shadow (or
counterpart)' of the traditional Mathematics over fields.

In particular, an idempotent version of Interval Analysis can
be constructed
\cite{24}. The idempotent interval arithmetics appear to be remarkably
simpler than their traditional analog. For example, in the traditional
interval arithmetic multiplication of intervals is not distributive with
respect to interval addition, while idempotent interval arithmetics
conserve distributivity. Idempotent interval arithmetics are useful
for reliable computing.

\section{ Idempotent linearity}

Let $S$ be a commutative idempotent semiring.

The following example of a noncommutative idempotent semiring is
very important.

{\sc Example 3.} Let $\Mat_n(S)$ be a set of all
$S$-valued matrices, i.e. coefficients of these matrices are elements
of $S$. Define the sum $\oplus$ of matrices $A = \|a_{ij}\|$,
$B = \|b_{ij}\| \in \Mat_n(S)$ as $A \oplus B = \|a_{ij} \oplus b_{ij}\|
\in \Mat_n(S)$. The {\it product} of two matrices $A \in \Mat_n(S)$ and
$B \in \Mat_n(S)$ is a matrix $AB \in \Mat_n(S)$ such that $AB =
\|\bigoplus_{k = 1}^m a_{ik} \odot b_{kj}\|$. The set $\Mat_n(S)$
of square matrices is an idempotent semiring with respect to these
operations. If $\0$
is the zero of $S$, then the matrix $O = \|o_{ij}\|$, where $o_{ij} = \0$,
is the zero of $\Mat_n(S)$; if $\1$ is the unity of $S$,
then the matrix $E = \|\delta_{ij}\|$, where $\delta_{ij} = \1$ if $i = j$
and $\delta_{ij} = \0$ otherwise, is the unity of $\Mat_n(S)$.

Now we discuss  an idempotent analog of a linear space. A set $V$  is
called a {\it semimodule} over $S$ (or an $S$-semimodule) if it
is equipped with an idempotent commutative associative addition operation
$\oplus_V$ and a multiplication $\odot_V$:~$S \times V \to V$
satisfying the
following conditions: for all $\lambda$, $\mu \in S$, $v$, $w \in V$
\begin{itemize}
\item $(\lambda \odot \mu) \odot_V v = \lambda \odot_V (\mu \odot_V v)$;
\item $\lambda \odot_V (v \oplus_V w)
= (\lambda \odot_V v) \oplus_V (\lambda \odot_V w)$;
\item $(\lambda \oplus \mu) \odot_V v
= (\lambda \odot_V v) \oplus_V (\mu \odot_V v)$.
\end{itemize}
An $S$-semimodule $V$ is called a {\it semimodule with zero} if
${\0} \in S$ and there exists a {\it zero} element
${\0}_V \in V$ such that for all $v \in V$, $\lambda \in S$
\begin{itemize}
\item ${\0}_V \oplus_V v = v$;
\item $\lambda \odot_V {\0}_V = {\0} \odot_V v = {\0}_V$.
\end{itemize}

{\sc Example 4. Finitely generated free semimodule.} The simplest
$S$-semimodule is the direct product $S^n = \{\, (a_1, \ldots, a_n) \mid
a_j \in S, j = 1, \ldots, n \,\}$. The set of all endomorphisms $S^n \to
S^n$ coincides with the semiring $\Mat_n(S)$ of all $S$-valued matrices
(see example~3).

The theory of $S$-valued matrices is similar to the well-known
Perron--Fro\-be\-ni\-us theory of nonnegative matrices, well advanced
and has very many applications, see, e.g., \cite{1}, \cite{3},
\cite{5}--\cite{15}, \cite{17}, \cite{21}, \cite{24}, \cite{29},
\cite{30}--\cite{32}).

\medskip

{\sc Example 5. Function spaces.} An {\it idempotent function space}
${\cal{F}}(X;S)$ consists of functional defined on a set $X$ and
taking their values in an idempotent semiring $S$. It is a subset
 of the set of all maps $X \to S$ such that
if $f(x)$, $g(x) \in {\cal{F}}(X;S)$ and $c \in S$, then $(f
\oplus g)(x) = f(x) \oplus g(x) \in {\cal{F}}(X;S)$
and $(c \odot f)(x) = c \odot f(x) \in
{\cal{F}}(X;S)$; in other words, an idempotent function space is another
example of an $S$-semimodule. If the semiring $S$ contains
a zero element
$\0$ and ${\cal{F}}(X;S)$ contains the zero constant function $o(x) =
\0$, then the function space ${\cal{F}}(X;S)$ has the structure of a
semimodule with zero $o(x)$ over the semiring $S$. If the set $X$
is finite
we get the previous example.

Recall that the idempotent addition defines a standard partial order in
$S$. An important example of an idempotent functional space is the space
${\cal{B}}(X;S)$ of all functions $X \to S$ bounded from above with
respect to the partial order $\pcle$ in $S$. There are many interesting
spaces of this type including ${\cal{C}}(X;S)$ (a space of continuous
functions defined on a topological space $X$), analogs of the Sobolev
spaces, etc (see,
e.g., \cite{17}, \cite{21}, \cite{23}, \cite{28}--\cite{30}
for details).

According to the correspondence principle, many important concepts, ideas
and results can be converted from usual Functional Analysis to Idempotent
Analysis. For example, an idempotent scalar product in ${\cal{B}}(X;S)$
can be defined by the formula
$$
 \langle\varphi,\psi\rangle = \int_X^\oplus \varphi(x) \odot \psi(x)\, dx,
$$
where the integral is defined as the `$\sup$' operation (see example 1).

\medskip

{\sc Example 6. Integral operators.} It is natural to construct
idempotent analogs of integral operators of the form
$$
K:\, \varphi(y) \mapsto (K\varphi)(x)
   = \int_Y^\oplus K(x,y) \odot \varphi(y)\, dy,
$$
where $\varphi(y)$ is an element of a functional space ${\cal{F}}_1(Y;S)$,
$(K\varphi)(x)$ belongs to a space ${\cal{F}}_2(X;S)$ and $K(x,y)$ is a
function $X \times Y \to S$. Such operators are {\it linear}, i.e. they are
homomorphisms of the corresponding functional semimodules. If $S = \rmax$,
then this definition turns into the formula
$$
(K\varphi)(x) = \sup_{y \in Y}\, (K(x,y) + \varphi(y)).
$$
Formulas of this type are standard for optimization problems.

\section{Superposition principle}

\quad\  In Quantum Mechanics the superposition principle means that the
Schr\"odi\-n\-ger equation (which is basic for the theory) is linear.
Similarly in Idempotent Mathematics the idempotent superposition
principle means that some important and basic problems and equations
(e.g., optimization problems, the Bellman equation and its versions
and generalizations, the Hamilton-Jacobi equation) nonlinear in the
usual sense can be  treated as linear over appropriate idempotent
semirings, see \cite{26}--\cite{30}, \cite{17}.

The linearity of the Hamilton-Jacobi equation over $\rset_{\min}$
(and $\rset_{\max}$) can be deduced from the usual linearity (over
$\Bbb{C}$) of the corresponding Schr\"odinger equation by means of the
dequantization procedure described above (in Section 4). In this case
the parameter $h$ of this dequantization coincides with $i\hbar$ ,
where $\hbar$ is the Planck constant; so in this case $\hbar$ must take
imaginary values (because $h>0$; see \cite{23} for details). Of
course, this is closely related to variational principles of
mechanics.

The situation is similar for the differential Bellman equation, see
\cite{17}.

It is well-known that discrete versions of the Bellman equation can be
treated as linear over appropriate idempotent semirings. The
so-called {\it generalized stationary} (finite dimensional)
{\it Bellman equation} has the form
$$
   X = AX \oplus B,
$$
where $X$, $A$, $B$ are matrices with elements from an idempotent
semiring and the corresponding matrix operations are described in
example 3 above; the matrices $A$ and $B$ are given (specified)
and it is necessary to determine $X$ from the equation.

B.A. Carr\'e \cite{5} used the idempotent linear algebra to
show that different optimization problems for finite graphs can be
formulated in a unified manner and reduced to solving these
Bellman equations, i.e., systems of linear algebraic equations over
idempotent semirings. For example, Bellman's method of solving
shortest path problems corresponds to a version of the Jacobi method
for solving systems of linear equations, whereas Ford's algorithm
corresponds to a version of the Gauss-Seidel method.

\section{Correspondence principle for computations}

Of course, the idempotent correspondence principle is valid for
algorithms as well as for their software and hardware implementations
\cite{20}--\cite{22}. Thus:

\begin{quote}
{\it If we have an important and interesting numerical algorithm, then
there is a good chance that its semiring analogs are important and
interesting as well.}
\end{quote}

In particular, according to the superposition principle,
analogs of linear
algebra algorithms are especially important. Note that
numerical algorithms
for standard infinite-dimensional linear problems over idempotent
semirings (i.e., for
problems related to idempotent integration, integral operators and
transformations, the Hamilton-Jacobi and generalized Bellman equations)
deal with the corresponding finite-dimensional (or finite) `linear
approximations'. Nonlinear algorithms often can be approximated by linear
ones. Thus the idempotent linear algebra is a basis for the idempotent
numerical analysis.

Moreover, it is well-known that linear algebra algorithms are convenient
for parallel computations; their idempotent analogs admit
parallelization as
well. Thus we obtain a systematic way of applying parallel computation to
optimization problems.

Basic algorithms of linear algebra (such as inner product of two vectors,
matrix addition and multiplication, etc.) often do not depend on
concrete semirings, as well as on the nature of domains containing the
elements of vectors and matrices. Algorithms to construct the closure
$A^*={\1}\oplus A\oplus A^2\oplus\cdots\oplus A^n\oplus\cdots=
\bigoplus^{\infty}_{n=1} A^n$ of an idempotent matrix $A$ can be derived
from standard methods for calculating $({\1} -A)^{-1}$. For the
Gauss--Jordan
elimination method (via LU-decomposition) this trick was used in \cite{31},
and the corresponding algorithm is universal and can be applied both to
the Bellman equation and to computing the inverse of a real (or complex)
matrix $({\1} - A)$. Computation of $A^{-1}$ can be derived
from this universal
algorithm with some obvious cosmetic transformations.

 Thus it seems reasonable to develop universal algorithms that can deal
equally well with initial data of different domains sharing the same
basic structure \cite{21}, \cite{22}.

\section{Correspondence principle for hardware design}

A systematic application of the correspondence principle to computer
calculations leads to a unifying approach to software and hardware
design.

The most important and standard numerical algorithms have many hardware
realizations in the form of technical devices or special processors.
{\it These devices often can be used as prototypes for new hardware
units generated by substitution of the usual arithmetic operations
for its semiring analogs and by addition tools for performing neutral
elements $\0$ and} $\1$ (the latter usually is not difficult). Of course,
the case of numerical semirings consisting of real numbers (maybe except
neutral elements) is the most simple and natural \cite{20}--\cite{22}.
Note that for semifields (including $\Bbb R_{\max}$ and $\Bbb R_{\min}$)
the operation of division is also defined.

Good and efficient technical ideas and decisions can be transposed
from prototypes into new hardware units. Thus the correspondence
principle generated a regular heuristic method for hardware design.
Note that to get a patent it is necessary to present the so-called
`invention formula', that is to indicate a prototype for the suggested
device and the difference between these devices.

Consider (as a typical example) the most popular and important algorithm
of computing the scalar product of two vectors:
\begin{equation}
(x,y)=x_1y_1+x_2y_2+\cdots + x_ny_n.
\end{equation}
The universal version of (1) for any semiring $A$ is obvious:
\begin{equation}
(x,y)=(x_1\odot y_1)\oplus(x_2\odot y_2)\oplus\cdots\oplus
(x_n\odot y_n).
\end{equation}
In the case $A=\Bbb R_{\max}$ this formula turns into the following one:
\begin{equation}
(x,y)=\max\{ x_1+y_1,x_2+y_2, \cdots, x_n+y_n\}.
\end{equation}

This calculation is standard for many optimization algorithms, so
it is useful to construct a hardware unit for computing (3). There
are many different devices (and patents) for computing (1) and every
such device can be used as a prototype to construct a new device for
computing (3) and even (2). Many processors for matrix multiplication
and for other algorithms of linear algebra are based on computing
scalar products and on the corresponding `elementary' devices
respectively, etc.

There are some methods to make these new devices more universal than
their prototypes. There is a modest collection of possible operations
for standard numerical semirings: max, min, and the usual arithmetic
operations. So, it is easy to construct programmable hardware
processors with variable basic operations. Using modern technologies
it is possible to construct cheap special-purpose multi-processor
chips implementing examined algorithms. The so-called
systolic processors are
especially convenient for this purpose. A systolic array is a
`homogeneous' computing medium consisting of elementary
processors, where the general scheme and processor connections
are simple and regular. Every elementary processor pumps data in and
out performing elementary operations in a such way that the
corresponding data flow is kept up in the computing medium; there
is an analogy with the blood circulation and this is a reason for the
term `systolic', see e.g. \cite{18}, \cite{31}.

Of course, hardware implementations for important and popular basic
algorithms can increase the speed of data processing.

\section{ Correspondence principle for software design}

Software implementations for universal semiring algorithms are not
so efficient as hardware ones (with respect to the computation speed)
but are much more flexible. Program modules can deal with abstract (and
variable) operations and data types. Concrete values for these
operations and data types can be defined by the corresponding
input data. In this case concrete operations and data types are generated
by means of additional program modules. For programs written in
this manner it is convenient to use a special techniques of the
so-called object oriented (and functional) design, see, e.g.,
\cite{25}, \cite{16}. Fortunately, powerful tools supporting the
object-oriented software design have recently appeared including compilers
for real and convenient programming languages (e.g. $C^{++}$ and Java).

There is a project to obtain an implementation of the correspondence
principle approach to scientific calculations in the form of a
powerful software system based on a collection of universal
algorithms. This approach ensures a working time reduction for
programmers and users because of the software unification.
The arbitrary
necessary accuracy and safety of numeric calculations can be ensured
as well.

The system contains several levels (including programmer and
user levels) and many modules.

Roughly speaking, it is divided into three parts. The first part
contains modules that implement domain
modules (finite representations of
basic mathematical objects). The second part implements universal
(invariant) calculation methods. The third part contains modules
implementing model dependent algorithms. These modules may be
used in user programs written in $C^{++}$ and Java.

\centerline{The following modules and algorithms
implementations are in
progress:}
\medskip
\centerline{Domain modules:}
\smallskip
infinite precision integers;

rational numbers;

finite precision rational numbers;

finite precision complex rational numbers;

fixed- and floating-slash rational numbers;

complex rational numbers;

arbitrary precision floating-point real numbers;

arbitrary precision complex numbers;

p-adic numbers;

interval numbers;

ring of polynomials over different rings;

idempotent semirings $R(\max, \min)$,
$R(\max, +)$, $R(\min, +)$, interval idem-

potent semirings

and others.

\centerline{Algorithms:}

linear algebra;

numerical integration;

roots of polynomials;

spline interpolations and approximations;

rational and polynomial interpolations and approximations;

special functions calculation;

differential equations;

optimization and optimal control;

idempotent functional analysis

and others.

This software system may be especially useful for designers
of algorithms, software engineers, students and mathematicians.

International Sophus Lie Centre

Nagornaya, 27--4--72, Moscow 113186 Russia

E-mail: litvinov@islc.msk.su

\end{document}